\documentclass[fleqn,usenatbib]{mnras}

\usepackage{newtxtext,newtxmath}

\usepackage[T1]{fontenc}

\DeclareRobustCommand{\VAN}[3]{#2}
\let\VANthebibliography\thebibliography
\def\thebibliography{\DeclareRobustCommand{\VAN}[3]{##3}\VANthebibliography}

\usepackage{graphicx}	
\usepackage{amsmath}	

\usepackage{xcolor}

\title[Direct contact binary planetesimal formation]{Direct contact binary planetesimal formation from gravitational collapse}

\author[J. T. Barnes et al.]{
Jackson T. Barnes,$^{1}$\thanks{E-mail: barne383@msu.edu}
Stephen R. Schwartz$^{2,3}$
Seth A. Jacobson,$^{1}$
\\
$^{1}$Department of Earth and Environmental Sciences, Michigan State University, 288 Farm Lane, East Lansing, MI 48824, United States\\
$^{2}$Planetary Science Institute, 1700 East Fort Lowell, Suite 106, Tucson, AZ 85719-2395, United States\\
$^{3}$Instituto de F\'isica Aplicada a las Ciencias y la Tecnolog\'ias, Universidad de Alicante, San Vicent del Raspeig, Alicante E-03690, Spain
}

\date{Accepted XXX. Received YYY; in original form ZZZ}

\pubyear{\the\year{}}

\begin{document}
\label{firstpage}
\pagerange{\pageref{firstpage}--\pageref{lastpage}}
\maketitle

\begin{abstract}
Bilobate contact binaries comprise a significant fraction of the relict Kuiper Belt, which includes the exemplary contact binary (486958) Arrokoth.
The surfaces of its lobes contain similar amounts of highly volatile chemical species and few craters, indicating formation in a homogeneous and gentle environment.
Arrokoth's bilobate shape was initially hypothesized to have formed via the direct gravitational collapse of a pebble cloud in the solar system's protoplanetary disk.
However, alternative hypotheses have proposed that Arrokoth may be the result of binary planetesimal formation and the subsequent dynamical evolution of the binary components into contact through external perturbations over long timescales.
Here, we show that contact binary planetesimals like Arrokoth can form directly from the gravitational collapse of pebble clouds.
We used a soft-sphere discrete element method (SSDEM) to discover that planetesimals form a wide variety of shapes, including bilobate contact binaries.
This method creates planetesimals as particle-aggregates with particles resting upon each other's surfaces via mutual surface penetration.
The formation of contact binaries in our simulations strengthens the hypothesis that Arrokoth, and perhaps many other contact binaries in the Kuiper Belt, formed directly as bilobate objects from gravitational collapse, and so their shapes and surfaces record the era of planet formation.
\end{abstract}




\section{Introduction}

Bilobate contact binaries exist throughout the solar system, including in the asteroidal \citep{Fujiwara2006,Benner2006,Magri2011,Zegmott2021,Lopez-Oquendo2022,Carry2023,Levison2024}, cometary \citep{Britt2004,Thomas2013,Massironi2015,Jorda2016}, and trans-Neptunian populations \citep{Sheppard2002,Sheppard2004,Lacerda2014,Buie2019,Stern2019,Thirouin2019a,Rabinowitz2020,Ashton2023}.
Many contact binary asteroids and comets are likely second-or-later-generation objects created after catastrophic collisional disruptions \citep[e.g.,][]{Michel2013,Schwartz2018,Bottke2023}, or as consequences of rotation-induced internal failure \citep[e.g.,][]{Jacobson2011,Margot2015,Wimarsson2024}, volatile outgassing \citep[e.g.,][]{Hirabayashi2016}, or of subcatastrophic reshaping collisions \citep[e.g.,][]{Jutzi2017a,Jutzi2017b}. Contact binaries in the trans-Neptunian population, however, must have formed \textit{in situ} and did not evolve due to significant collisional, rotational, or surface processing evolutions \citep{Davidsson2016,Thirouin2017b,Thirouin2019b,Steckloff2021,Simon2024}.
It is estimated that contact binaries comprise a significant fraction of the Plutino population \citep[40--50\%;][]{Thirouin2018} and the pristine and dynamically cold classical Kuiper Belt population \citep[10--25\%;][]{Sheppard2004,Lacerda2011,Thirouin2018,Thirouin2019a}.
It is also estimated that approximately 85\% of contact binaries may not be recognized in observations due to unfavorable object orientations \citep{Lacerda2011}.
We note that the amplitude of an object's photometric light curve cannot uniquely distinguish between contact binaries, tight binaries, or elongated objects \citep[e.g.,][]{Harris2020}, so these estimates possess a difficult to quantify systematic uncertainty.

The cold classical Kuiper Belt object (486958) Arrokoth is a relict planetesimal and a quintessential contact binary   \citep{Buie2019,Stern2019}.
Arrokoth's orbit is sufficiently far from the giant planets to avoid significant dynamical evolution \citep{Porter2018}, generally distant from other small bodies resulting in only limited collisional evolution \citep{Greenstreet2019,Mao2021}, and too distant from the Sun for sublimative or radiative torques to have significant consequences \citep{McKinnon2020,Steckloff2021}.
The best estimates of its shape were made from the 2019 flyby of NASA's New Horizons spacecraft \citep{Spencer2020,Keane2022,Porter2024}, as shown in the central panel of Fig.~\ref{fig:cb_composite}.
The surfaces of its small and large lobes, Wenu and Weeyo, lack strong surface albedo and color differences \citep{Stern2019,Grundy2020}, contain similar amounts of highly volatile chemical species \citep{Grundy2020,Lisse2021}, and only a modest amount of craters with a similar inferred crater age \citep{Spencer2020,Schenk2021}.
From these observations, the lobes must have a common origin and formed in a gentle environment that did not sublimate volatile ices or cause significant surface deformation \citep{Stern2019,McKinnon2020,Spencer2020,Marohnic2021}.
Furthermore, Arrokoth's smaller lobe, Wenu, has a hexagonal-like shape that may be due to the aggregation of multi-km sized planetesimals at low $\lesssim$\,1\,m\,s$^{-1}$ velocities \citep{Stern2023}.
Observations from the 2019 New Horizons flyby led to the immediate inference that Arrokoth formed via a gravitational collapse process \citep{Stern2019}.

The gravitational collapse of clouds of mm-size pebbles directly produces self-gravitating (i.e., diameter $\gg$\,1\,km) planetesimals, bypassing intermediary sizes associated with significant growth barriers \citep[for reviews, see][]{Johansen2014,Simon2024}.
Gravitational collapse may occur as a consequence of the streaming instability, a runaway process that creates gravitationally unstable clouds of pebbles due to momentum feedback from inward drifting solids onto slower orbiting gas in a protoplanetary disk \citep{Youdin2005b,Youdin2007b,Johansen2007b}.
Within these collapsing clouds, pebbles bounce off of one another or fragment such that kinetic energy in the cloud dissipates as gravitational potential energy is released due to cloud contraction.
Collisional dissipation also drives a decrease in particle relative velocities resulting in gentler and more frequent collisions---the conditions for effective sticking and accretion and ultimately growth to planetesimal size-scales \citep{Johansen2007b,Nesvorny2010,WahlbergJansson2014,WahlbergJansson2017}.
As the pebble cloud gravitationally collapses, its mass is concentrated into a rapidly decreasing volume, and by the conservation of angular momentum, its rotation rate increases.
Since the cloud cannot contract into a single object rotating faster than the critical spin break-up limit \citep[e.g.,][]{Scheeres2007}, the cloud contracts to form a central near-equal-mass binary or multi-component system as well as possible additional ejected single or multi-component systems \citep{Nesvorny2010,Nesvorny2019a,Robinson2020,Nesvorny2021,Barnes2025}.
Gravitational collapse does not release enough energy to significantly sublimate volatile ices \citep{Bierson2019}, and collision speeds are relatively low, limiting surface deformation \citep{WahlbergJansson2014,Blum2017,WahlbergJansson2017,Fulle2017}.
Simulated binary systems formed from gravitational collapse are consistent with observations of relict Kuiper Belt binaries \citep{Nesvorny2010,Nesvorny2019a,Fraser2017,Barnes2025}, and gravitational collapse can create oblate (i.e., flattened) objects \citep{Lorek2024,Barnes2025}.
However, until now, no model yet demonstrated that bilobate contact binaries can be directly produced as an outcome of gravitational collapse.

Since the discovery of Arrokoth's bilobate shape, the precise process that formed the two lobes, Wenu and Weeyo, has been debated.
There was immediate speculation that the lobes were brought together directly by the gravitational collapse process itself \citep{Stern2019}.
Yet, since the gravitational collapse process has only been demonstrated as-of-yet to create separated binary systems \citep{Nesvorny2010,Nesvorny2019a,Robinson2020,Nesvorny2021,Barnes2025}, other hypotheses suggested that the two lobes were first stable components of a binary that only later were brought together by subsequent mutual dynamical evolution.
Proposed dynamical evolution to drive the components into contact include in-spiraling due to gas drag \citep{McKinnon2020}, Kozai-Lidov oscillations \citep{Grishin2020}, gas drag assisted by Kozai-Lidov oscillations \citep{Lyra2021}, or a sophisticated dynamical interplay between Kozai-Lidov oscillations, tidal friction, and perturbations from the giant planets \citep{Brunini2023}.
Thus, whether Arrokoth's components were driven together by dynamics within the collapsing cloud or if they instead required subsequent processes remains unresolved.

\section{Methods}
Using a soft-sphere discrete element method (SSDEM), we were able to successfully model the gravitational collapse of clouds of particles with realistic densities to form planetesimal systems with multiple components, diverse shapes, and varied spin states \citep{Barnes2025}.
Specifically, our simulations used the PKDGRAV \textit{N}-body integrator \citep{Dikaiakos1996,Richardson2000,Stadel2001} and its soft-sphere discrete element method \citep[SSDEM;][]{Schwartz2012,Zhang2017} (ver. 08/03/2018), which monitors short-distance particle interactions with contact forces as well as long-distance gravitational interactions \citep{Richardson2000,Schwartz2012,Zhang2017}.
Critically, the SSDEM does not perfectly merge particles upon collision, instead particles interact with one another other through mutual surface penetration and therefore simulate long-lasting particle contacts via parameterized contact physics \citep{Cundall1978}.
The PKDGRAV SSDEM models normal and tangential particle contact forces as spring and dashpot systems, which enables the simulation of restitutional and frictional forces \citep{Schwartz2012,Zhang2017}.
Thus, the SSDEM enables the creation of planetesimals as particle-aggregates with distinct shapes and spin states \citep{Barnes2025}.
Furthermore, because particles do not perfectly merge with one another in the SSDEM, planetesimals can experience both accretion and decretion \citep[see Fig.~1 in][]{Barnes2025}, and thus an SSDEM paints a more complex depiction of planetesimal formation.

The PKDGRAV SSDEM N-body simulations were designed to simulate the collapse of a cloud with the same mass as an approximately $100$\,km-sized planetesimal \citep{Barnes2025}.
A pebble cloud would have likely contained approximately a septillion ($10^{24}$) mm-sized pebbles \citep{Johansen2015}, but such a high resolution N-body simulation is effectively impossible, so similar to prior works \citep{Nesvorny2010,Nesvorny2019a,Robinson2020,Nesvorny2021}, we simulated a system of $10^5$ super-particles, each with a radius of about 2 km.
If a spherical pebble cloud initially extended uniformly over an appreciable fraction of the Hill volume associated with its mass and the cloud was initially rotating about its center of mass at or just below its associated circular (i.e., Keplerian) velocity, then it has already been shown that the trans-Neptunian wide binary population can be well reproduced by numerical N-body simulations \citep{Nesvorny2010,Nesvorny2019a,Robinson2020,Nesvorny2021}.
Using 54 PKDGRAV SSDEM N-body simulations \citep{Barnes2025}, we reproduced this result that was first achieved via perfect merging N-body methods \citep{Nesvorny2010}.
Of these 54, we ran 30 simulations varying the initial cloud rotation rate, and with an additional 24 simulations, we adjusted the parameters controlling the SSDEM contact physics and found the results to be insensitive to those adjustments.
For further details regarding the simulation set-up and the first results from these PKDGRAV SSDEM simulations, please see \citet{Barnes2025}.

\section{Results}

\begin{figure*}
    \centering
    \includegraphics[width=2\columnwidth]{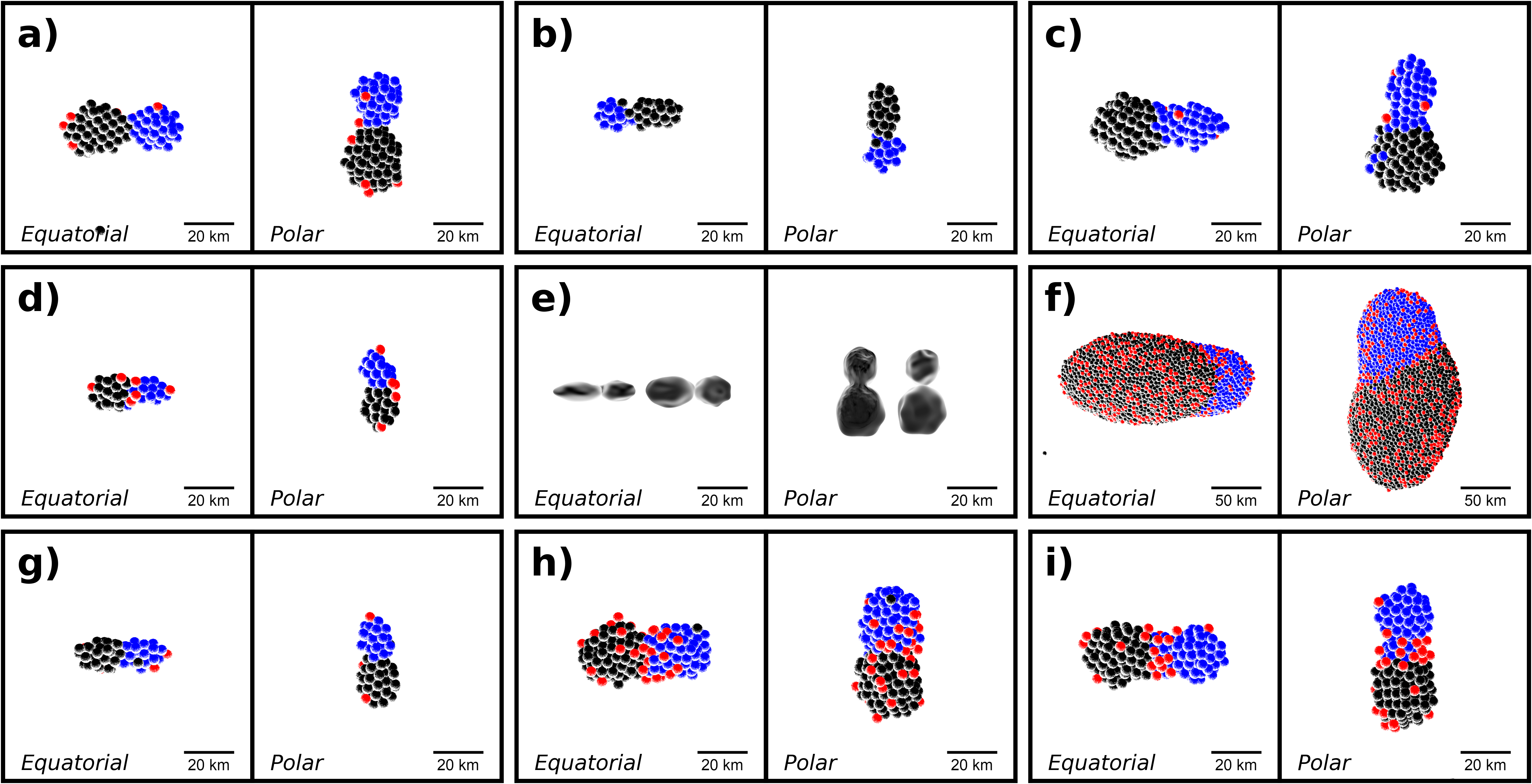}
    \caption{\textbf{Several examples of contact binary planetesimals created using the PKDGRAV SSDEM (panels a--d and f--i) as well as two shape models of (486958) Arrokoth from \citet{Keane2022} (panel e, left) and \citet{Porter2024} (panel e, right).} 
    Each contact binary is shown as a pair of images from two perspectives, equatorial and polar, with associated scale bars included for reference.
    The black-colored lobes have the most mass prior to impact and are therefore considered the primary lobes, while the blue-colored lobes have less mass and so are the secondary lobes.
    Red-colored particles accreted onto their respective contact binary planetesimal after the lobes made contact.
    }
    \label{fig:cb_composite}
\end{figure*}

Out of the $54$ PKDGRAV SSDEM N-body simulations, approximately $3$\% of planetesimals form as contact binaries, see Fig.~\ref{fig:cb_composite} for examples.
In detail, $29$ contact binary planetesimals were identified from a population of $834$ planetesimals, with each of these particle-aggregates containing $45$ or more particles (encompassing approximately the same volume as a spherical planetesimal with a diameter of about $15$\,km)---planetesimals below this size threshold were too poorly resolved to exhibit a distinguishable bilobate shape.
To distinguish contact binaries from all other objects, each candidate contact binary planetesimal was required to have had two mutually orbiting lobes that made contact, and after impact, the resulting planetesimal continued to possess a distinct bilobate shape by eye.
Of these twenty-nine simulated contact binaries, twenty-four have a very clearly defined bilobate shape and five were considered to be borderline cases with a less pronounced neck.

\begin{figure}
    \centering
    \includegraphics[width=\columnwidth]{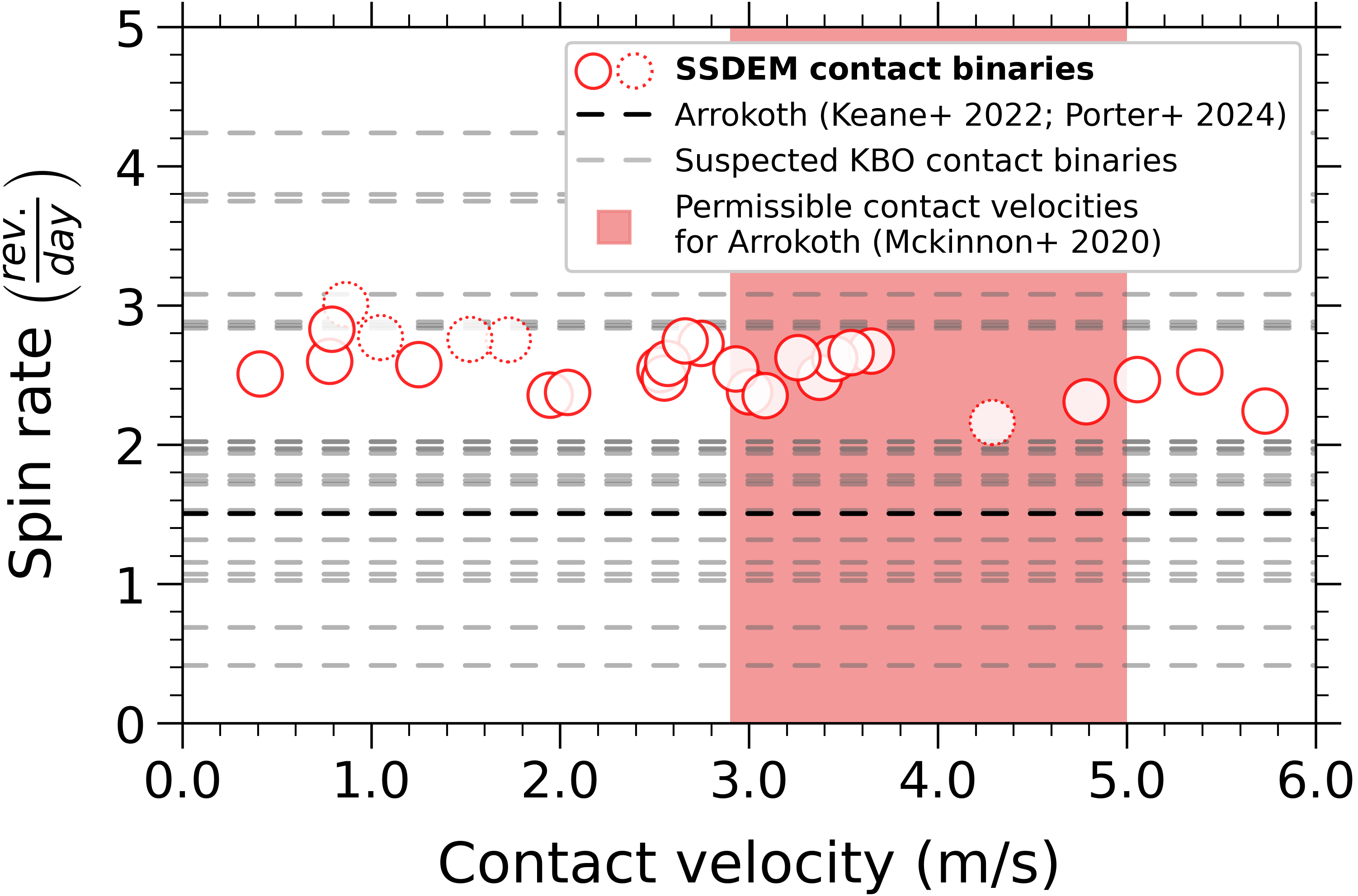}
    \caption{\textbf{Contact binary spin rates from simulated and observed populations as a function of the collision velocities of the mutually orbiting lobes.}
    SSDEM contact binaries are shown as red solid and dotted circles.
    The dotted circles are simulated planetesimals that did not have as clear a bilobate shape as indicated by red-solid circles, but may be considered contact binaries because they are each the result of two mutually orbiting planetesimals that made contact to create a near-bilobate elongated shape.
    The spin rates of suspected contact binaries from the Kuiper Belt are displayed as gray dashed lines \citep{Kern2006,Spencer2006,Lacerda2011,Thirouin2017a,Thirouin2017b,Thirouin2018,Thirouin2019a,Rabinowitz2020,Thirouin2022,Thirouin2024,Porter2025}, and the spin rate of the confirmed contact binary Arrokoth is displayed as a black dashed line \citep{Buie2020}.
    One value of contact velocity, for the planetesimal in Fig.~\ref{fig:cb_composite}f, resides outside the current axes; it has a contact velocity of $\sim$\,16.9\,m\,s$^{-1}$ with a spin rate of 2.65\,rev/day.
    Theorized permissible contact velocities for the Arrokoth lobes are shown as a red vertical band \citep{McKinnon2020}.}
    \label{fig:contact_vel}
\end{figure}

All of the identified contact binaries began as two separate aggregate objects that collided together after first existing as a gravitationally bound binary.
These binary planetesimals are created together, growing in a mutual gravitational potential and accreting mostly unbound particles and particle aggregates.
In one case, the contact binary is one of the central objects of the overall collapsing cloud, but in all other cases, the contact binary forms from a binary pair that is ultimately ejected from the collapsing cloud; the central objects of most clouds only accrete a fraction of the total initial mass of the collapsing cloud \citep{Nesvorny2010,Robinson2020,Nesvorny2021,Barnes2025}.
This is consistent with the observation that Arrokoth is smaller \citep[about 20\,km;][]{Keane2022,Porter2024} than the approximate 100\,km size-scale objects that are theoretically predicted to dominant the mass distribution of collapsing pebble clouds \citep{Johansen2007b,Simon2024}.

At late times during the collapse, bound binaries continue to interact with other planetesimals in the collapsing cloud such that their semi-major axes decreases with time, even while their masses, shapes, and spin rates remain effectively constant.
The mutual orbital energy shared between the binary members is lost via energy exchange with other passing planetesimals, which typically have relative kinetic energies less than the absolute binding energy of the binary---an application of Heggie-Hill's Theorem that `hard binaries harden' \citep{Heggie1975,Hills1975}.
In the case of contact binaries, this inspiralling ultimately leads to a collision between the two components.

In general, the collision is very gentle, and, as shown in Fig.~\ref{fig:contact_vel}, all SSDEM contact binaries except one made contact at velocities ranging between 0.4--5.8\,m\,s$^{-1}$---the outlier made contact at a much higher 16.9\,m\,s$^{-1}$.
Many simulated contact binaries make contact within the range of velocities, 2.9--5.0\,m\,s$^{-1}$, hypothesized from geophysical and geomorphological considerations for the collision of Arrokoth's two lobes \citep{McKinnon2020}.
Thus the PKDGRAV SSDEM simulations independently validate that analysis, confirming that contact binary formation directly from gravitational collapse occurs when contact velocities are below about 6\,m\,s$^{-1}$.

\begin{figure}
    \centering
    \includegraphics[width=\columnwidth]{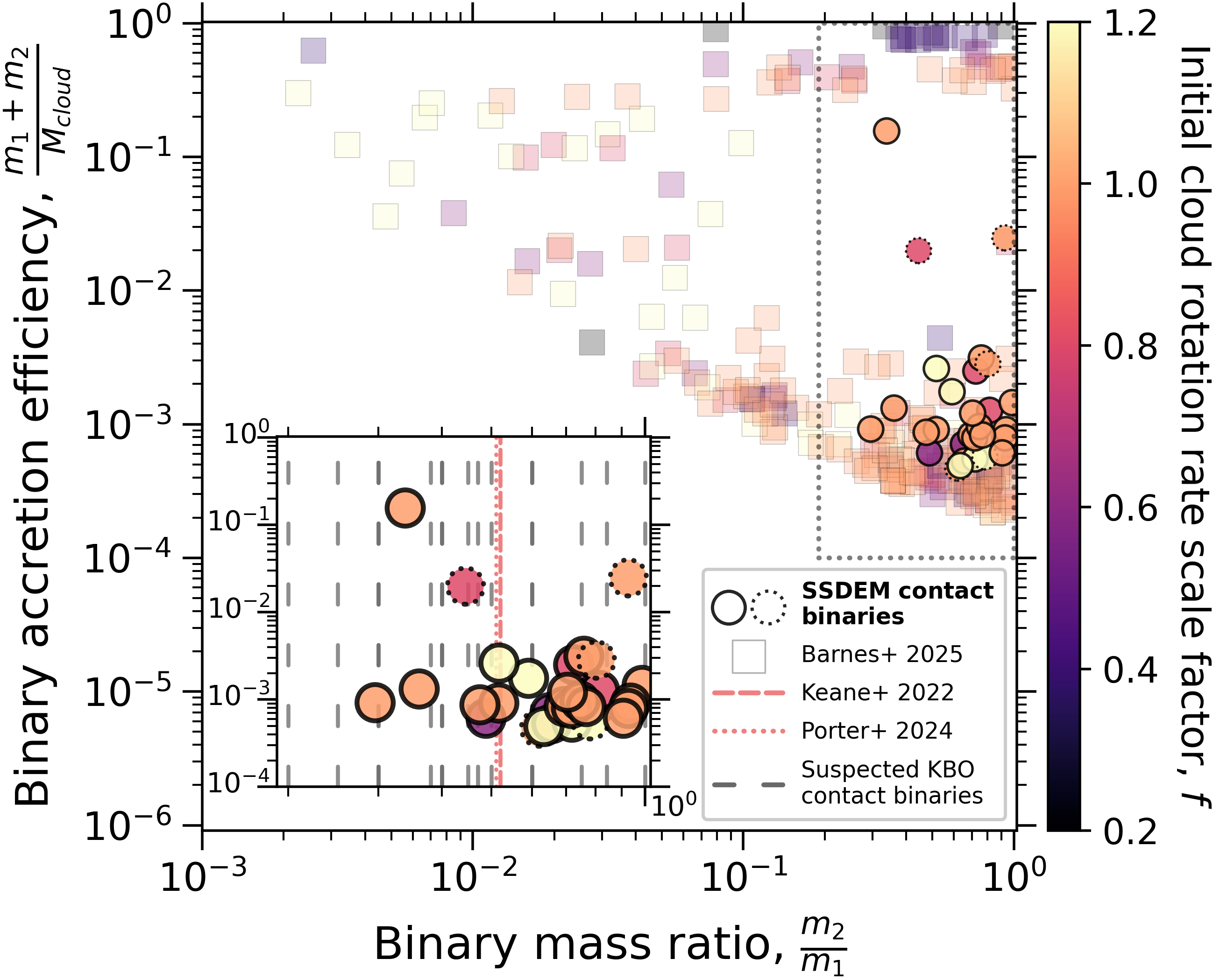}
    \caption{\textbf{Contact binary lobe accretion efficiencies $\left(m_1+m_2
    \right)/M_{cloud}$, where $M_{cloud}$ is the initial pebble cloud mass, are shown as a function of their lobe mass ratios $m_2/m_1$.}
    The main figure shows the full scope of mass ratios and accretion efficiencies of binary planetesimal systems from the SSDEM simulations.
    The gray dotted line indicates the region displayed in the left sub-figure, which is a comparison of all simulated and observed contact binary planetesimals.
    SSDEM contact binaries are displayed as circles (solid and dotted), SSDEM binary planetesimal systems \citep{Barnes2025} are displayed as squares, the mass ratios of suspected contact binaries from the Kuiper Belt are displayed as gray vertically dashed lines, \citep{Lacerda2007,Thirouin2017a,Thirouin2017b,Thirouin2018,Thirouin2019a,Farkas-Takacs2020,Rabinowitz2020,Thirouin2022,Thirouin2024}, and the mass ratios of the confirmed contact binary Arrokoth are displayed as pink vertically dashed \citep{Keane2022} and dotted \citep{Porter2024} lines.
    As in Fig.~\ref{fig:contact_vel}, the dotted circles are simulated planetesimals that did not have as clear a bilobate shape, but may be considered contact binaries because they are each the result of two mutually orbiting planetesimals that made contact to create a near-bilobate elongated shape.
    }
    \label{fig:mass_ratios}
\end{figure}

Contact binary planetesimals created by the PKDGRAV SSDEM may indicate a primordial preference for slower rotation rates removed from the spin break-up limit, which is at about 3.6\, rev/day (6.6\,hr rotation period) for equal-lobe-mass contact binaries with bulk densities of about 1\,g\,cm$^{-3}$.
The simulated contact binaries exhibit a narrow range of post-contact spin rates typically between 2.1--3.0\,rev/day (8.0--11.5\,hr rotation periods), which are comparable to observed candidate contact binaries as shown in Fig.~\ref{fig:contact_vel}.
The majority of our modeled contact binaries exhibit rotation periods moderately faster than Arrokoth, which rotates at $\sim$\,1.51\,rev/day (15.93\,hr).
Additionally, approximately 62\% of simulated contact binaries have prograde rotations.
This primordial prograde preference should be compared to future observations as rotation pole information for Kuiper Belt contact binaries becomes determined.

The relatively limited collisional evolution in the cold classical region of the Kuiper Belt may nevertheless be responsible for some of the more slowly rotating observed contact binaries relative to those from the PKDGRAV simulations.
Although the sublimation of surface volatiles likely does not dramatically alter its bulk shape (estimated surface changes of 10--100 m) or spin rate \citep{Steckloff2021}, cratering collisions may have gradually slowed Arrokoth to its current state via a prolonged collisional history \citep{McKinnon2022}.
This hypothesis assumes that a decrease in spin is viable based on recently modeled values for crushing strength \citep{Housen2018} and bulk porosity \citep[$\sim$50\%;][]{Grundy2020}, which would allow the impact craters on Arrokoth's surface to form via largely inelastic collisions that result in surface compaction as opposed to the displacement of the underlying target material.
Ultimately, Arrokoth's rotation period may have been reduced from a period comparable to the simulated contact binaries as a result of collisions with hundreds of approximately km-sized Kuiper Belt objects \citep{McKinnon2022}.
These considerations may explain the minor discrepancies between the spin periods of the simulated and observed contact binaries.

Contact binaries have equal or near-equal sized lobes by definition, so their formation requires pre-existing similar sized components.
In the PKDGRAV SSDEM simulations, equal-mass lobes are the products of super-particle clouds with initial rotation rate scale factors $f\sim$~0.6--1.2, whereas near-equal mass lobes result from clouds with slower rotation rates $f<$~0.6, as shown in Fig.~\ref{fig:mass_ratios}.
A similar relationship was previously established for simulated binary systems with accretion efficiencies $10^{-4}<(m_1+m_2)/M_{cloud}<10^{-2}$ \citep{Barnes2025}.

Contact binaries do not form as the largest planetesimals within the collapsing cloud but rather as ejected cloud components during the collapse.
All of the simulated contact binaries have low accretion efficiencies, as shown in Fig.~\ref{fig:mass_ratios} and indicating that most of the collapsing cloud mass is incorporated into other bodies.
The contact binaries are similar to a larger population of binary systems created in the collapsing cloud but no longer bound to the more efficiently accreting central planetesimal system \citep{Robinson2020,Barnes2025}.

\begin{figure}
    \centering
    \includegraphics[width=\columnwidth]{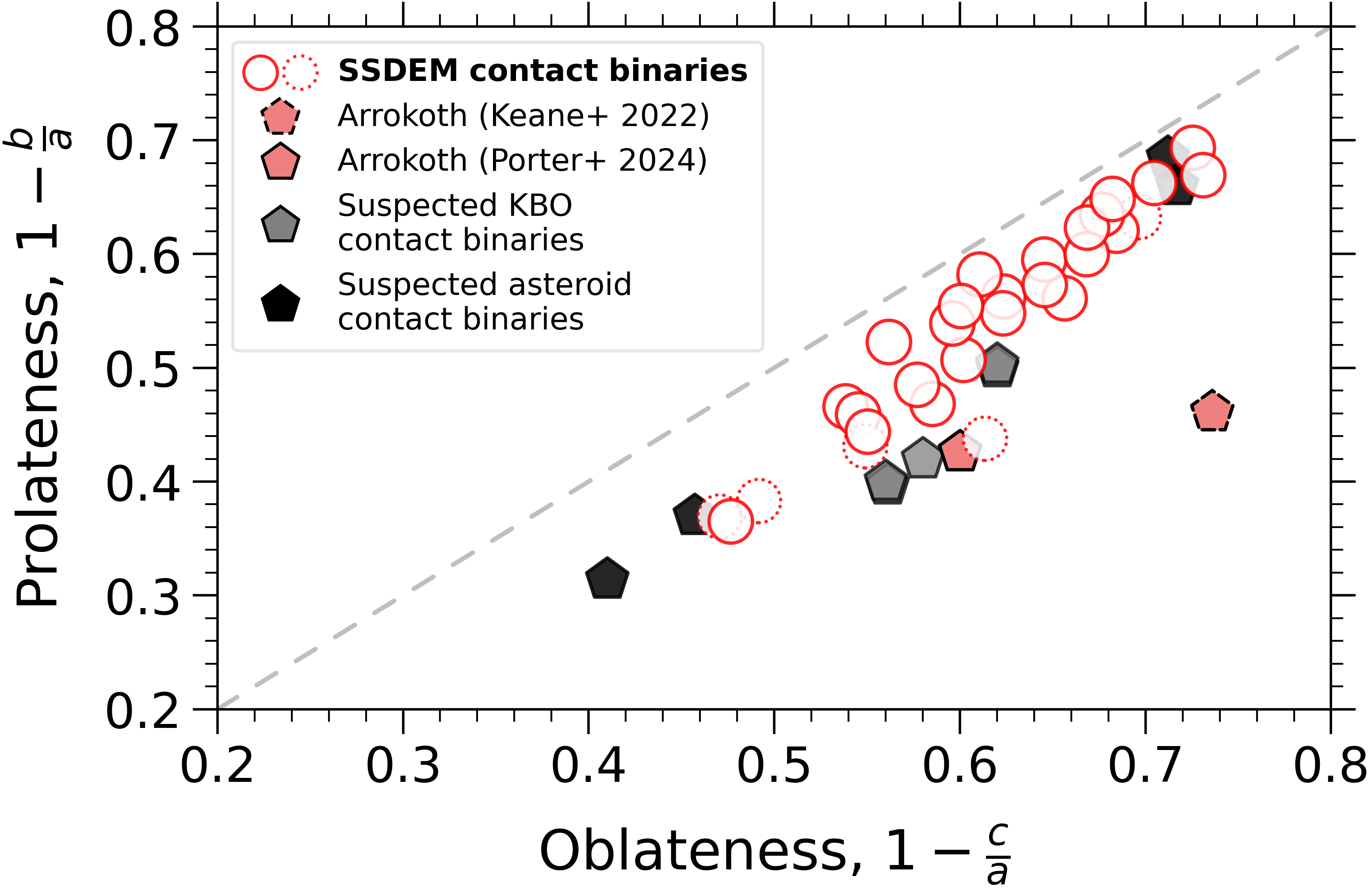}
    \caption{\textbf{The prolateness $\left(1 - b/a\right)$ and oblateness $\left(1 - c/a\right)$ of simulated and observed contact binary planetesimals.}
    Simulated contact binary planetesimals are shown as solid and dotted red circles as in Figs.~\ref{fig:contact_vel} and~\ref{fig:mass_ratios}.
    As in Fig.~\ref{fig:contact_vel}, the red-dotted circles are simulated planetesimals that did not have as clear a bilobate shape as indicated by red-solid circles, but may be considered contact binaries because they are each the result of two mutually orbiting planetesimals that made contact to create a near-bilobate elongated shape.
    SSDEM contact binaries are compared to suspected contact binaries in the asteroid belt (201 Penelope, \citet{Shepard2015}; 216 Kleopatra, \citet{Shepard2018}, 413 Edburga, \citet{Shepard2015}) and Trojan (624 Hektor, \citet{Descamps2015}) populations (black pentagons), suspected contact binaries in the Kuiper Belt (2004 TT357, \citet{Thirouin2017a}; 2014 JL80, 2014 JO80, 2014 JQ80, \citet{Thirouin2018}; and 2004 VC131, \citet{Thirouin2019a}; gray pentagons), and two models of 486958 Arrokoth from \citet{Keane2022} (red pentagon) and \citet{Porter2024PDS,Porter2024} (red pentagon with dashed line).
    }
    \label{fig:cb_bulk_semi_axes}
\end{figure}

The PKDGRAV SSDEM simulated contact binaries have prolate shapes with moderately rounded lobes.
As shown in Fig.~\ref{fig:cb_bulk_semi_axes}, each of the modeled contact binaries can be characterized by their prolateness $\left(1 - \frac{b}{a}\right)$ ranging from $\sim$0.37--0.69 and oblateness $\left(1 - \frac{c}{a}\right)$ ranging from $\sim$0.47--0.73.
The prolateness and oblateness are defined as $\left(1 - \frac{b}{a}\right)$ and $\left(1 - \frac{c}{a}\right)$, respectfully, using the $a$, $b$, and $c$ semi-axes of the body so that $a \geq b \geq c$. 
For observed systems, the semi-axes are estimated from the derived figure that best matches observations, and for the PKDGRAV simulated systems, they are derived from each body's principal moments assuming a constant density of 1~g~cm$^{-3}$ such that $a \geq b \geq c$.
The majority of our SSDEM contact binaries exhibit axes ratios similar to the limited sample of observed primordial contact binaries throughout the solar system, including those from the Kuiper Belt \citep{Descamps2015,Shepard2015,Thirouin2017a,Thirouin2018,Shepard2018,Thirouin2019a} and two distinct models of Arrokoth \citep{Keane2022,Porter2024PDS,Porter2024}.

No simulated contact binary matches the distinct flattened shape first estimated for Arrokoth \citep{Keane2022}.
Critically, because Arrokoth's larger lobe, Wenu, has a markedly flatter shape than its smaller and rounder lobe, Weeyo \citep[especially for][]{Keane2022}, it skews Arrokoth's bulk shape toward greater values of oblateness $1 - \frac{c}{a}$, far removed from the simulated and other observed contact binaries.
It was initially suggested that some other process \citep[e.g., surface and near-subsurface volatile sublimation][]{Zhao2021} may be responsible for modifying Arrokoth's shape post-contact.
Alternatively, it has also been argued that sublimation is not effective at modifying Arrokoth's bulk shape but rather only its surface.
Its shape may therefore be entirely primordial, and its surface may be more-evolved \citep{Steckloff2021}.
An updated shape model of Arrokoth from \citet{Porter2024} has noticeably rounder lobes and is therefore better matched by the simulated contact binaries, not requiring such a significant shape evolution \citep{Steckloff2021}.

Arrokoth's larger lobe, Wenu, is characterized by a near-hexagonal profile about its equator 
\citep[see Fig.~\ref{fig:cb_composite};][]{Stern2019,Porter2024}.
This distinct shape may be owed to its formation by the gentle aggregation of an assortment of approximately 5-km-sized mounds originating from a common source \citep{Stern2023}.
Some SSDEM contact binaries experienced a similar accretion process.
One specific case, the SSDEM contact binary most similar to Arrokoth (Fig.~\ref{fig:cb_composite}\hyperref[fig:cb_composite]{a}), was formed by the aggregation of several smaller objects with $D_{eq}\sim$~8--15~km and supplemented by numerous individual particles with $D_{eq}\sim$~4~km.
The larger sizes of these mounds are owed to our fundamental resolution limit, whereby our collapsing clouds were initialized with a monodisperse distribution of $\sim$2-km radius particles.
If we were to increase the resolution beyond our current limits or use a polydisperse assortment of particles, it may be possible to create mounds with sizes more commensurate with mounds formed in recent simulations \citep{Stern2023}.

Generally, these results demonstrate that pristine contact binaries, like Arrokoth, can be directly produced by the gravitational collapse of a pebble cloud.
Indeed, note the particular similarity between Fig.~\ref{fig:cb_composite}\hyperref[fig:cb_composite]{a} and Fig.~\ref{fig:cb_composite}\hyperref[fig:cb_composite]{c} with Arrokoth (Fig.~\ref{fig:cb_composite}\hyperref[fig:cb_composite]{e}).
Other contact binaries from throughout the solar system are likely not primordial and their lobes were brought into contact via other mechanisms, however their shapes and the shapes of the simulated contact binaries are similar.
For instance, Fig.~\ref{fig:cb_composite}\hyperref[fig:cb_composite]{b} with asteroid 1996 HW1, Fig.~\ref{fig:cb_composite}\hyperref[fig:cb_composite]{d} to asteroid Itokawa, Fig.~\ref{fig:cb_composite}\hyperref[fig:cb_composite]{f} to asteroid 1999 JV6, Fig.~\ref{fig:cb_composite}\hyperref[fig:cb_composite]{g} to asteroid Toutatis, and Fig.~\ref{fig:cb_composite}\hyperref[fig:cb_composite]{i} with comet 8P/Tuttle.
This also suggests a similarity of the contact velocities and geometries between the simulated and observed systems, despite the differing dynamical mechanisms that likely brought them into contact.

Gravitational collapse also creates contact binaries as components of multi-component planetesimal systems.
In the PKDGRAV SSDEM simulations, four contact binaries have orbiting satellites: three contact binaries have one satellite and one contact binary has two satellites. 
Additionally, two contact binaries are the satellites of multi-component systems with more than two gravitationally bound components.
These two types of contact-binary-multi-component systems are similar to the primordial trans-Neptunian binary system Lempo-Hiisi and the more-evolved binary asteroid system Dinkinesh-Selam, respectively, although the latter likely formed through a different mechanism.

\section{Summary \& Conclusions}
Contact binaries can form directly from the gravitational collapse of pebble clouds.
We used the PKDGRAV N-body integrator with the SSDEM to model the gravitational collapse of a cloud of 10$^{5}$~2-km radius super-particles---analogous to collapsing pebble clouds which are hypothesized to contain $\sim10^{24}$~mm-sized solids \citep{Johansen2015}.
These simulations created contact binaries with bulk and lobe shapes as well as lobe mass ratios characteristic of those observed throughout the solar system, including the cold classical Kuiper Belt object, Arrokoth, the quintessential example of a pristine relict contact binary planetesimal.
Simulated contact binary lobes collided at $\leq$\,6\,m\,s$^{-1}$ consistent with hypothesized contact velocities for Arrokoth's lobes \citep{McKinnon2020}.
Simulated contact binaries have a spin rate of approximately 2.1-3.0\,rev/day which similar to observed contact binaries are well below the critical spin limit.
Future work would benefit from higher resolution models and a polydisperse distribution of particles, which may allow for populations of modeled contact binaries with more diverse shapes, sizes (i.e., below our current by-eye viewing limit), and increased detail on the surface of each lobe.

\section*{Acknowledgments}
SAJ and JTB acknowledge direct support from the National Science Foundation (NSF) (award number: AST-2406891).
SAJ and JTB also acknowledge support for JTB from the NASA Michigan Space Grant Consortium Graduate Fellowship (award number: 80NSSC20M0124).
SAJ and JTB was supported by use of the Extreme Science and Engineering Discovery Environment (XSEDE), which was supported by National Science Foundation (NSF) (award numbers: ACI-1053575 and ACI-1548562). 
This work was also supported in part through computational resources and services provided by the Institute for Cyber-Enabled Research at Michigan State University.

\section*{Data Availability}

The processed data from our PKDGRAV simulations and data used to directly generate the figures of our contact binaries will be available on the Zenodo page (doi:10.5281/zenodo.17872208) associated with this article.



\bibliographystyle{mnras}
\bibliography{biblio}








\bsp	
\label{lastpage}
\end{document}